\documentclass[12pt,english]{article}
\usepackage[T1]{fontenc}
\usepackage[latin9]{inputenc}
\usepackage[letterpaper]{geometry}
\usepackage{amsmath}
\usepackage{graphicx}

\makeatletter
\newcommand{\lyxaddress}[1]{
	\par {\raggedright #1
	\vspace{1.4em}
	\noindent\par}
}

\usepackage{times}

\makeatother

\usepackage{babel}
\usepackage[sorting=none, maxbibnames=99]{biblatex}
\begin{document}
\global\long\def\ub{\boldsymbol{u}}%

\global\long\def\gradu{\nabla\ub}%

\global\long\def\cb{\boldsymbol{c}}%

\global\long\def\sb{\boldsymbol{\sigma}}%

\global\long\def\db{\boldsymbol{\delta}}%

\global\long\def\gdot{\dot{\gamma}}%

\global\long\def\wi{\text{Wi}}%

\title{Pathological Rheology of Non-Stretching Entangled Polymers: Finite-Time Blow-Up Predictions}
\author{Vickie Chen, Brandon Wang, and Joseph D. Peterson}
\maketitle

\lyxaddress{Department of Chemical and Biomolecular Engineering, University of
California, Los Angeles, 420 Westwood Plaza, Los Angeles CA 90095}
\begin{abstract}
The non-stretching approximation of polymer rheology simplifies a
constitutive equation but fundamentally changes its behavior in fast
flows, and the circumstances under which fast flows emerge cannot
always be predicted a-priori. In this paper, we consider two simple
flows for which shear rates are bounded in the original RP model but
diverge to infinity in finite time for the non-stretching RP model.
The disparity between the full and non-stretching models can be resolved
by extending the non-stretching approximation to second order in accuracy. 
\end{abstract}

\section{\protect\label{sec:Introduction}Introduction}

Entangled polymers are an important material for modern life, with
hundreds of billions of pounds produced every year. When designing,
optimizing, and troubleshooting industrial polymer processing equipment,
theoretical and computational tools can play a critical role in predicting
and interpreting unexpected flow behaviors. With clever approximations,
abstractions, and simplifications, it is often possible to make theoretical/conceptual
tools more generalizable and computational/numerical tools faster
and more efficient. For every simplification, however, there is always
a risk of inadvertently stripping a theory or model of some crucial
functionality.

For virtually every computational approach to studying entangled polymers
in complex flows, detailed simulations become more difficult as the
polymers become increasingly entangled. This is true for particle-based
simulations \cite{premvzoe2003particle,muller2003particle,kindt2007single},
discrete slip link models \cite{schieber2003full,khaliullin2010application,andreev2013approximations},
and even continuum theories \cite{likhtman2000microscopic,graham2003microscopic,read2008entangled}.
In all cases, the escalation of computational complexity is due to
an increase in numerical ``stiffness'', where increasing chain length
leads to an ever-widening gap between the fastest and slowest relaxing
degrees of freedom in the system. To ameliorate stiffness concerns,
one often presumes certain fast degrees of freedom to be ``equilibrated'':
small sections of a polymer can be represented as flexible Kuhn beads,
which are further coarse grained to slip springs, and so on until
coarse graining and closure approximations encounter structures that
cannot be considered equilibrated under the flow conditions of interest. 

In the limit of extremely large entanglement numbers, $Z\gg100$,
intra-tube Rouse modes - including those related to longitudinal stretch
relaxation - relax quickly compared to the slow processes of tube
renewal via reptation (curvilinear diffusion) and constraint release
\cite{doi1988theory}. For continuum models of entangled polymers,
it is therefore expedient to assume that polymers are (to leading
order) unstretched in flow, provided the rate of deformation remains
much slower than the slowest intra-tube Rouse relaxation process.
The non-stretching approximation has been used extensively in continuum
models of highly entangled polymers, from full-chain models like the
original Doi Edward model \cite{doi1988theory,likhtman2000microscopic}
to single-mode approximations like Rolie Poly (RP) \cite{likhtman2003simple,ianniruberto2001simple}.

The non-stretching approximation has proven very useful for understanding
``universal'' behaviors of entangled polymers under flow conditions
where the assumptions of the non-stretching approximation remain valid
\cite{adams2011transient,moorcroft2014shear}. Unfortunately, it is
not always possible to confirm a-priori whether a non-stretching approximation
is appropriate to a particular complex flow - i.e. whether the rate
of deformation will remain bounded on the time interval of interest
- prior to generating a solution for said flow. Examples of high shear
rates in start-up of slow shear flows are well documented in the so-called
``transient shear banding'' literature \cite{adams2009nonmonotonic,adams2011transient,moorcroft2014shear,carter2016shear,sharma2021onset}.

In gereral, a finite-time blow-up is not physically meaningful in
any model context; rather, it signals the omission of key physics
needed to regularize the solution. For example, infinite shear rates
could be prevented by introducing a small amount of viscosity or fluid
inertia \cite{sharma2021onset,moorcroft2014shear}. However, for physically
relevant results a non-stretching flow ought to be regularized by
re-introducing chain stretching. In our view, there are only two justifiable
approaches for reintroducing chain stretching, namely (a) reverting
back to the original (stiff) equations with chain stretch or (b) pursuing
higher order corrections to the non-stretching approximation. The
latter direction will be the principal focus of the present manuscript.

The organization of our work is as follows. In section \ref{sec:Equations-and-Nondimensionalizat},
we present the non-dimensionalized governing equations for the full
RP model, the conventional (first order) non-stretching approximation,
and a new second-order continuation that includes previously omitted
physics pertaining to chain stretching. In section \ref{sec:Examples-of-Finite-time},
we provide two definite examples of finite time blow-up in the non-stretching
RP model, demonstrating the need for regularization. In section \ref{sec:nRP2_success},
we show that the higher-order continuation of the non-stretching approximation
avoids blow-up and successfully follows the full RP model predictions.

\section{\protect\label{sec:Equations-and-Nondimensionalizat}Equations and
Nondimensionalization}

For well-entangled monodisperse linear polymers melts and solutions,
the Rolie Poly (RP) model \cite{likhtman2003simple} provides a simple
and effective means of predicting nonlinear rheology in both complex
and viscometric flows. The RP model accounts for stress relaxation
by reptation, chain retraction, and convective constraint release,
and generalizations of the RP model have further accounted for finite
extensibility \cite{kabanemi2009nonequilibrium}, polydispersity \cite{boudara2019nonlinear},
disentanglement dynamics \cite{dolata2023thermodynamically}, and
reversible scission reactions \cite{peterson2021predictions}. Here,
we present the non-dimensionalized governing equations of the RP model
with the regularized IM correction to ensure positive entropy production
\cite{ianniruberto2001simple}:

\begin{equation}
\overset{\nabla}{\cb}-(\boldsymbol{c}-\delta)-6Z(1-a)(\boldsymbol{c}+\beta a(\boldsymbol{c}-\db))\label{eq:full_RP_ndim}
\end{equation}

\begin{equation}
a=1/\lambda=\sqrt{\frac{3}{\text{tr}\cb}}
\end{equation}

\begin{equation}
\beta=\beta'\tanh(\Lambda Z(1-a))
\end{equation}

\begin{equation}
\sb=\cb-\db
\end{equation}

The configuration tensor $\cb$ describes changes in the second-moment
of the end-to-end vector of an entanglement segment, scaled to the
identity tensor $\db$ at equilibrium. Polymers are assumed to deform
affinely in flow, as indicated by the use of an upper convected Maxwell
derivative, $\overset{\nabla}{\cb}$, on the left hand side of equation
\ref{eq:full_RP_ndim}. The parameter $a=1/\lambda$ is the reciprocal
of the chain stretch parameter $\lambda$. The CCR parameter $\beta$
follows the regularized Ianniruberto correction, and to ensure that
the regularization error is below our tolerance for numerical integration
we $\Lambda=1000$ for all simulations.

The nondimensionalization scheme is as follows; the stress tensor
stress $\sb$ is scaled by a shear modulus $G_{e}$, time $t$ is
scaled by the reptation time $\tau_{D}$. For unidirectional shear
flows, distances are scaled by the ``gap dimension'' $H$, and the
velocity is scaled by $H/\tau_{D}$ for convenience.

For flows driven by a defined velocity $U_{c}$, the dimensionless
numbers that govern the RP model are the entanglement number $Z$,
and the Weissenberg number $\wi=U_{c}\tau_{D}/H$. For flows driven
by a defined shear stress $\sigma_{xy}$, the dimensionless shear
stress is a governing parameter that can be used in place of the Weissenberg
number.

When the entanglement number is very large, $Z\gg1$, stretch relaxation
occurs quickly compared to orientational relaxation. Provided the
rate of deformation remains slow compared to the rate of stretch relaxation,
$\wi\ll Z$, the dynamics of the RP model are, to leading order, independent
of the entanglement number, as described by the so-called ``non-stretching''
RP (nRP) model \cite{likhtman2003simple}:

\begin{equation}
\overset{\nabla}{\boldsymbol{c}}=-(\boldsymbol{c}-\delta)-\frac{2}{3}(\cb:\gradu)(\boldsymbol{c}+\beta(\boldsymbol{c}-\db))\label{eq:full_nRP_ndim}
\end{equation}

\begin{equation}
\text{tr}\cb=3
\end{equation}

\begin{equation}
\beta=\tanh\left(\frac{1}{9}\Lambda Z\cb:\gradu\right)
\end{equation}

\begin{equation}
\sb=\cb-\db
\end{equation}

If the non-stretching approximation is reframed as the leading order
solution to a regular perturbation expansion in the limit of $1/Z\ll1$,
i.e. $\cb\approx\cb_{0}+Z^{-1}\cb_{1}+\cdots$, a second-order continuation
of the nRP model, which we call nRP2, is given by:

\begin{equation}
\overset{\nabla}{\boldsymbol{c}}_{0}=-(\boldsymbol{c}_{0}-\boldsymbol{\delta})+6a_{1}(\boldsymbol{c}_{0}+\beta_{0}(\boldsymbol{c}_{0}-\boldsymbol{\delta}))
\end{equation}

\begin{equation}
\overset{\nabla}{\boldsymbol{c}}_{1}=-\boldsymbol{c}_{1}+6a_{2}(\boldsymbol{c}_{0}+\beta_{0}(\boldsymbol{c}_{0}-\boldsymbol{\delta}))+6a_{1}(\boldsymbol{c}_{1}+\beta_{1}(\boldsymbol{c}_{0}-\boldsymbol{\delta})+\beta_{0}(a_{1}\boldsymbol{c}_{0}+\boldsymbol{c}_{1}-a_{1}\boldsymbol{\delta}))
\end{equation}

\begin{equation}
a_{1}=-\frac{1}{9}\boldsymbol{c}_{0}:\nabla u\hspace{1cm}a_{2}=-\frac{1}{3}\left(\frac{\partial}{\partial t}a_{1}+\ub\cdot\nabla a_{1}\right)-\frac{1}{9}\gradu:\boldsymbol{c}_{1}+\left[-\frac{1}{3}+2\left(1+\beta_{0}\right)a_{1}\right]a_{1}
\end{equation}

\begin{equation}
\beta_{0}=-\beta'\tanh(\Lambda a_{1})\hspace{1cm}\beta_{1}=-\beta'\Lambda a_{2}\cosh^{-2}(\Lambda a_{1})
\end{equation}

\begin{equation}
\sigma=\boldsymbol{c}_{0}+\frac{1}{Z}\boldsymbol{c}_{1}-\boldsymbol{\delta}
\end{equation}

A full derivation and assessment of the nRP2 model will be provided
in a forthcoming manuscript \cite{chen2024derivation}. With the second-order
continuation, chain stretch is treated very differently; the leading
order terms implicitly describe a quasi-static relationship between
chain stretching and the instantaneous rate of deformation, but the
second-order terms show that chain stretching also depends on flow
history.

\section{\protect\label{sec:Examples-of-Finite-time}Pathologies of the nRP
model}

Here, we will provide two examples of finite-time blow-up in the nRP
model; nonlinear creep flow and startup steady Taylor-Couette flow.
In both cases, we will assume unidirectional flow starting from a
quiescent initial condition.

For the nonlinear creep flow, apply a fixed shear stress and observe
the resulting shear rate evolve over time. Above a critical shear
stress, steady flow solutions cease to exist and transiently the shear
rate blows-up in finite time. Here, finite time blow-up can be partly
expected based on the lack a valid steady state solution.

For the startup steady Taylor-Couette flow, however, valid steady
state solutions do exist - but a finite-time blow-up is encountered
during the flow transients en-route to steady state. These results
reinforce our claim in the introduction that it is generally not possible
to predict a-prior whether the nRP model will remain bounded in a
given flow.

\subsection{\protect\label{subsec:nonlinear_creep_nRP}Finite-time blow-up in
nonlinear creep test of nRP model}

We consider the RP and nRP model in steady simple shear flow between
two parallel plates separated by a distance $H$ with a constant dimensionless
shear stress $\sigma_{xy}$ applied at all times. The flow is assumed
to be unidirectional with a homogeneous strain rate at all times,
$\ub(t,x,y,z)=\gdot(t)y\boldsymbol{e}_{x}$.

Because we are ignoring inertia and viscous dissipation, the flow
response to a constant shear stress is divided into two stages. First,
there is an initial ``impulse'' deformation where the material instantaneously
deforms until it reaches the prescribed stress. This impulse deformation
is absorbed as a change in the initial condition. Second, following
the impulse deformation, the material begins to relax its stress and
must deform (or creep) with a shear rate $\gdot(t)$ to maintain the
prescribed shear stress.

\begin{figure}
\begin{centering}
\includegraphics[scale=0.25]{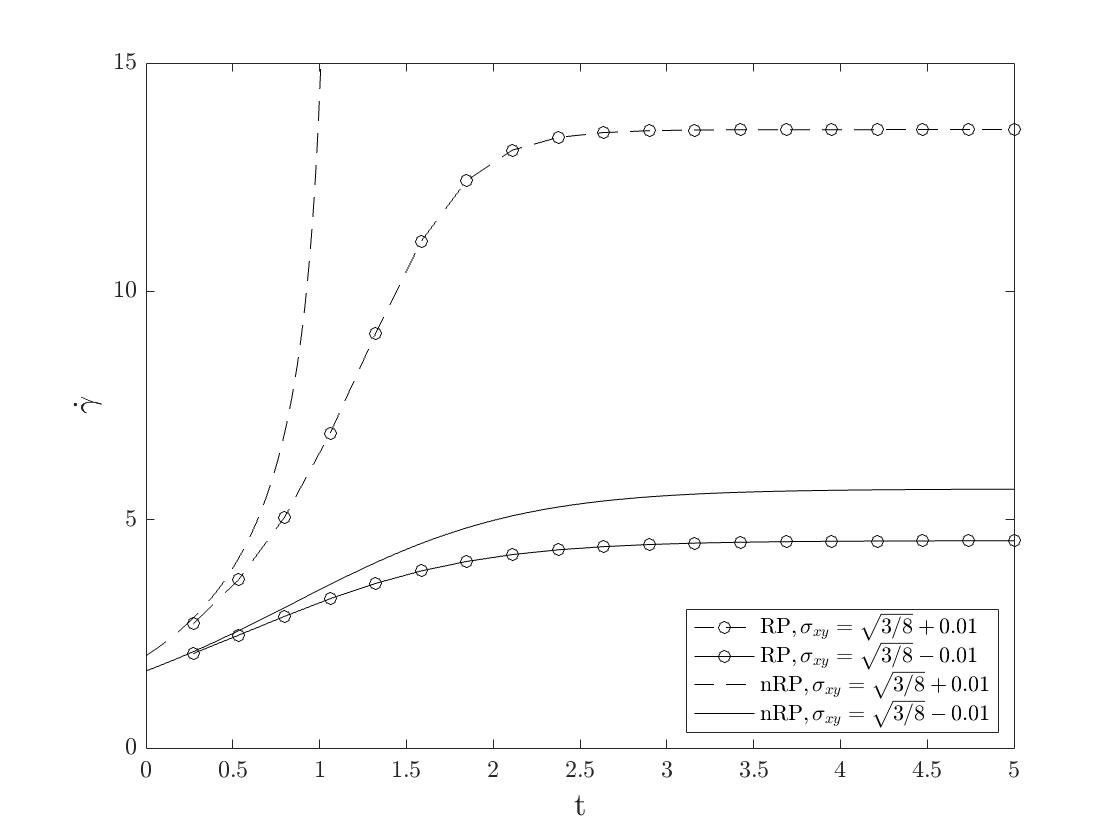}
\par\end{centering}
\caption{\protect\label{fig:nonlinear_creep}Nonlinear creep test for the nRP
model and the RP model with $Z=100$ and $\beta'=1$. The shear rate
is plotted over time for shear stresses $\sigma_{xy}=\sqrt{3/8}\pm0.01$.
For $\sigma_{xy}<\sqrt{3/8}$, the nRP model closely approximating
the RP model's behavior. For $\sigma_{xy}>\sqrt{3/8}$, the nRP model
exhibits a finite-time blow-up in the shear rate, while the RP model
stabilizes due to chain stretching.}
\end{figure}

In Figure \ref{fig:nonlinear_creep}, we present results for nonlinear
creep tests of the nRP model and the RP model with $Z=100$ and $\beta'=1$.
For shear stresses $\sigma_{xy}<\sqrt{3/8}$, a bounded solution for
$\gdot(t)$ exists at all times in both the RP and nRP models, and
the nRP model provides a good approximation of the full RP model behavior.
However, for $\sigma_{xy}>\sqrt{3/8}$, a steady flow solution no
longer exists in the nRP model and flow transients exhibit a finite-time
blow-up in the shear rate $\gdot(t)$. In all observed cases of finite-time
blow-up in nonlinear creep flow, we find that the the shear rate diverges
to infinity at a critical time $t^{*}$ as $\gdot(t)\sim(t-t^{*})^{-1/2}$.
Thus the total strain imposed $\gamma(t)=\int_{0}^{t}\gdot(t')dt'$
remains integrable.

Briefly, we summarize the mechanism of this finite-time blow-up. In
entangled polymers, deformation and chain retraction can cause an
over-rotation of tube segments; increasing the shear rate leads to
short-term increases in shear stress but contributes to over-rotation
in the long term. Attempting to maintain a shear stress that is unsustainable
(i.e. has no steady flow solution) leads to to progressively higher
and higher shear rates up to some critical strain where the over-rotation
has progressed to the point that an infinite shear rate is needed
to maintain the current shear stress. In the RP model, no finite-time
blow-up is observed; chain stretching becomes activated at sufficiently
high shear rates, offsetting the over-rotation effect and stabilizing
flow at a finite shear rate. 

\subsection{\protect\label{subsec:TC_nRP}Finite-time blow-up in Taylor-Couette
for nRP model}

The existence of a valid steady flow solution in the nRP model does
not guarantee the existence of a valid time-dependent trajectory to
reach that steady state. Here, we consider startup of unidirectional
shear flow in a Taylor Couette (concentric rotating cylinders) geometry.
Given an inner cylinder radius $R_{o}$ and outer cylinder radius
$R_{o}=R_{i}+H$, the outer cylinder remains fixed and the inner cylinder
rotates with velocity $U_{c}$. Once again, we assume a unidirectional
shear flow in the azimuthal direction, $\ub=u_{\theta}\boldsymbol{e}_{\theta}$
with a shear rate $\gdot=r\frac{\partial}{\partial r}\left[\frac{1}{r}\frac{\partial u_{\theta}}{\partial r}\right]$
that varies radially from the inner cylinder to the outer cylinder,
with the highest shear rates found at the inner cylinder wall.

\begin{figure}
\centering{}\includegraphics[scale=0.25]{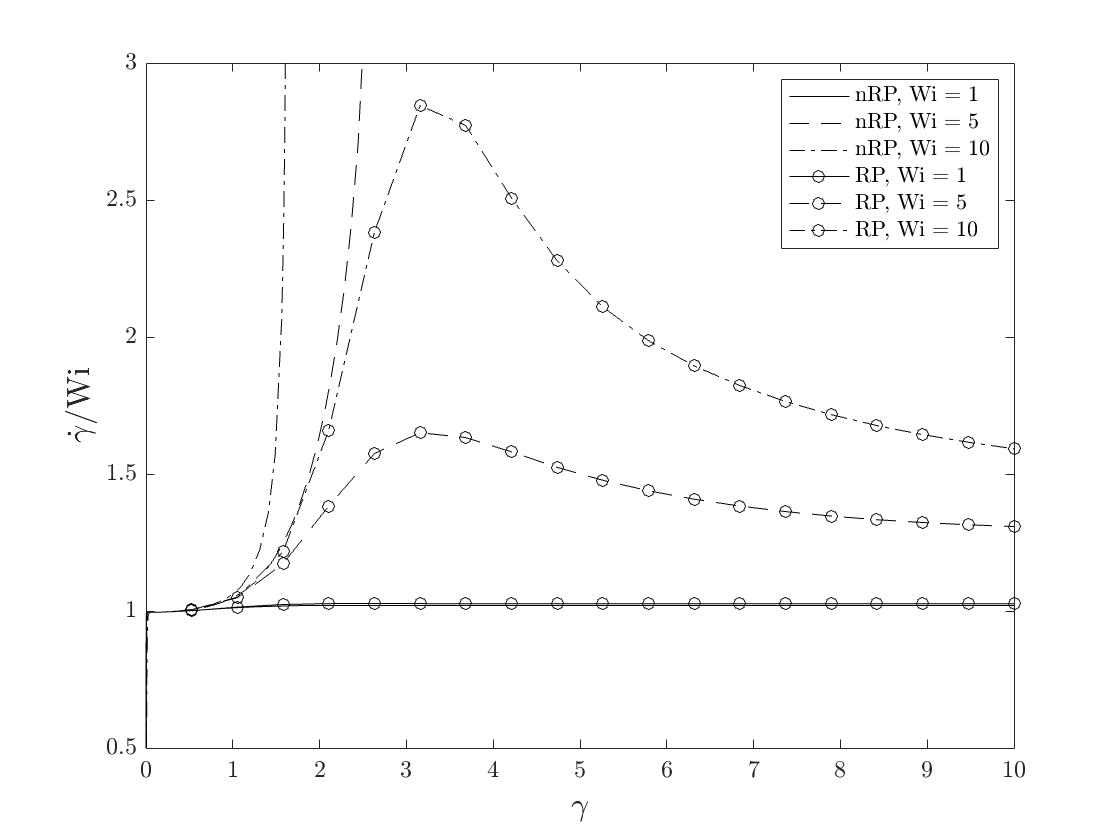} \caption{\protect\label{fig:nRP_TC}Comparison of shear rate at the inner wall
scaled by the Weissenberg number $\dot{\gamma}/\text{Wi}$, as a function
of the imposed strain for both the RP and nRP models in a Taylor-Couette
flow. The curvature parameter is $p=0.01$. The results are presented
for $Z=100$ and $\beta'=1$ with Weissenberg numbers Wi = 1,5 and
10. For Wi = 1, the nRP model agrees well with the RP model, but at
higher Weissenberg numbers, the nRP model predictions exhibit a blow-up
in the shear rate within finite time.}
\end{figure}

In Figure \ref{fig:nRP_TC}, we compare predictions for the RP model
with $Z=100$ against predictions of the nRP model for a Taylor-Couette
flow with curvature $p=1-R_{i}/R_{o}=0.01$. For a range of Wi$=1,5,10$
and 20, we compare the shear rate at the inner wall, $\gdot_{i}$,
scaled by the Weissenberg number Wi, as a function of the imposed
strain, $\gamma=\text{Wi}t$. For Wi$=1$, the nRP model predictions
show no pathological behavior and there is good agreement with the
RP model predictions. For higher Wi, however, the shear rate at the
inner cylinder wall $\gdot_{i}$ blows-up in finite time for the nRP
model, while no such issues are seen with the RP model predictions. 

\begin{figure}
\centering{}\includegraphics[scale=0.25]{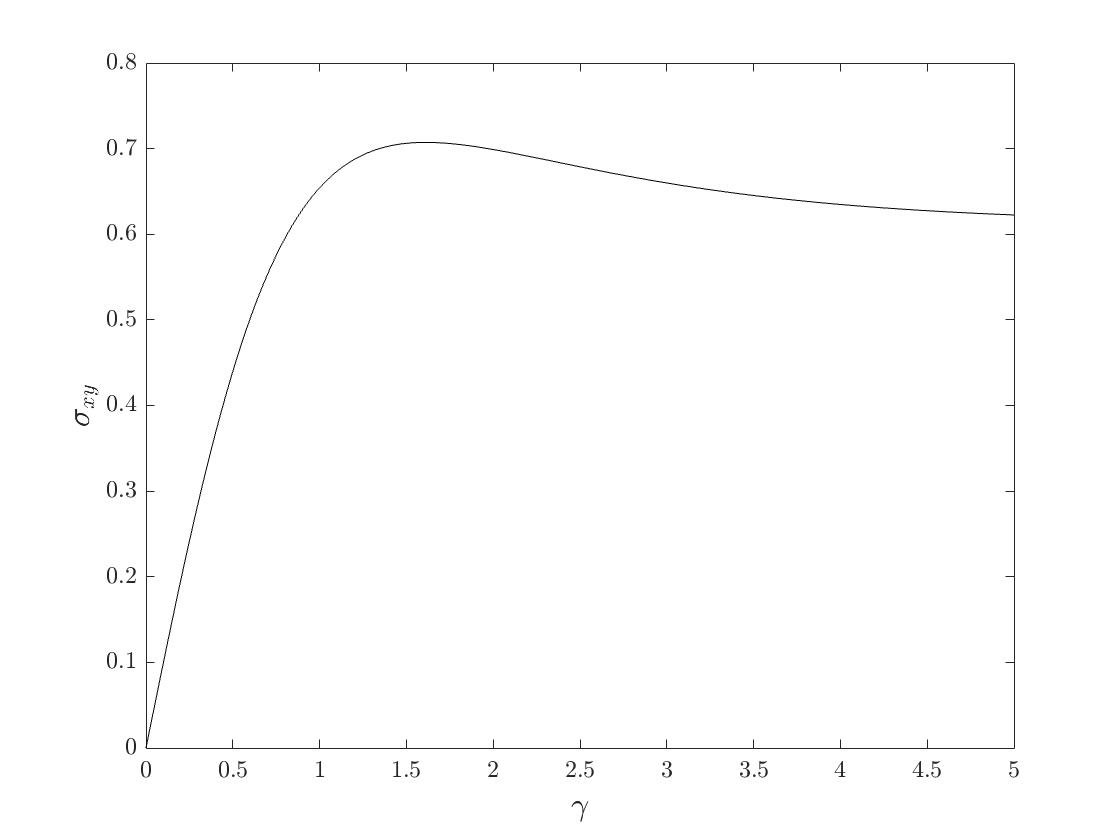}\caption{\protect\label{fig:stress_strain_nRP_highWi}Shear stress versus shear
strain for the nRP model in startup with $\protect\wi\gg1$. The maximum
shear stress occurs at $\gamma_{p}$\textasciitilde 1.5}
\end{figure}

Briefly, we summarize the mechanism of this finite-time blow-up. For
the nRP model at Wi$\gg1$, the reptation terms in equation \ref{eq:full_nRP_ndim}
can be ignored to leading order and the shear stress $\sigma_{xy}$
is purely a function of the applied strain $\gamma$ as shown in Figure
\ref{fig:stress_strain_nRP_highWi}. The maximum in shear stress is
associated with an over-rotation of chain segments, as discussed in
subsection \ref{subsec:nonlinear_creep_nRP}. If the stress/strain
curve of Figure \ref{fig:stress_strain_nRP_highWi} is valid for an
inertia free, unidirectional Taylor Couette flow context, the stress
overshoot creates a catastrophy as the material at the inner cylinder
approaches the apex of the stress strain curve - it is not possible
to satisfy momentum balance without having infinite shear rates concentrated
at the inner cylinder wall. Reintroducing a finite Wi breaks the deterministic
relationship between stress and strain, but there is still a positive
feedback problem; higher shear rates occur near the inner cylinder
wall to satisfy momentum balance, and higher shear rates produce an
increasingly deterministic stress/strain relationship, which in turn
favors higher shear rates near the inner cylinder walls and so on
until the shear rates at the inner cylinder wall becomes infinite
once again. In the RP model, this feedback cycle is halted when the
inner shear rate becomes so large that chains begin to stretch, breaking
the deterministic relationship between stress and strain.  

The relationship between stress overshoot and large shear rate gradients
has been previously explored in the ``transient shear banding''
literature \cite{moorcroft2013criteria,moorcroft2014shear}, but the
mathematical analysis therein was not rigorous \cite{sharma2021onset}
and gaps in the purported universality of the relationship have not
been well understood. Based on the discussion here, large gradients
in shear rate should not be expected for the Giesekus model \cite{giesekus1982simple}
because it lacks the separation of stretch/orientation timescales
that would give rise to a universal stress/strain curve for startup
shear flows in the limit of $1\ll\wi\ll Z$.

\section{\protect\label{sec:nRP2_success}Success of the nRP2 model}

\begin{figure}
\begin{centering}
(a) \includegraphics[scale=0.15]{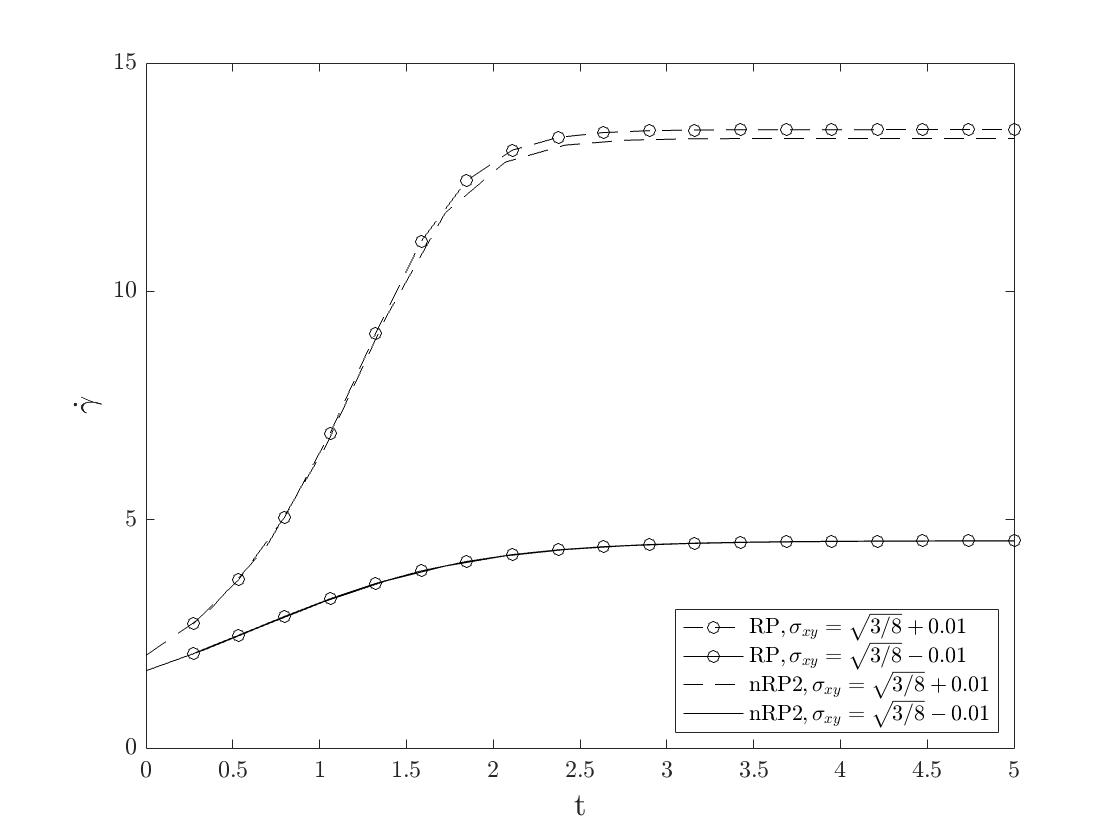}(b) \includegraphics[scale=0.15]{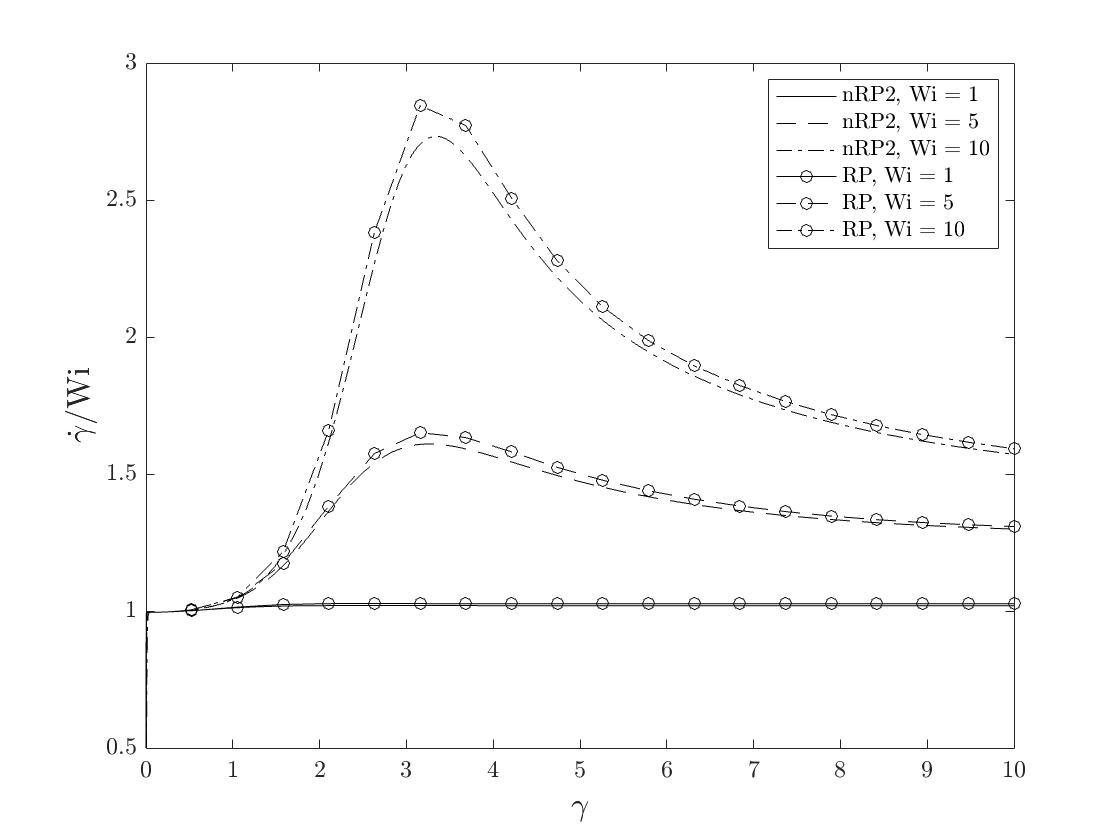}
\par\end{centering}
\caption{\protect\label{fig:nRP2_predictions}(a) Shear rate as a function
of time during nonlinear creep for RP and nRP2 models with Z=100 at
different stress levels $\sigma_{xy}=\sqrt{3/8}\pm0.01$. The results
of the nRP2 model aligns well with those of the RP without any finite-time
blow-up. (b) Scaled shear rate $\dot{\gamma}/\text{Wi}$ as a function
of strain for the RP and nRP2 models in a Taylor-Couette flowwith
Z=100 at Weissenberg numbers Wi=1,5,10. The predictions of the nRP2
model closely match those of the RP model, with no evidence of finite-time
blow-up.}

\end{figure}

In both of the finite-time blow-up scenarios that we explored in section
\ref{sec:Examples-of-Finite-time}, failure of the nRP model occured
due to the loss of chain stretching effects at high shear rates. By
contrast, the nRP2 model provides a more complete description of chain
stretching effects, including a history dependence of chain stretching
and its effect on chain retraction. Thus, it may be expected that
the nRP2 model can provide the benefits of a non-stretching approximation
without the liabilities of numerical stiffness. In Figure \ref{fig:nRP2_predictions},
we compare predictions of the RP and nRP2 model with $Z=100$ for
both nonlinear creep and start-up Taylor-Couette flow. Overall, we
find good agreement between the nRP2 and RP model predictions, with
no sign of a finite-time blow-up in either model.

\section{\protect\label{sec:Conclusions}Conclusions}

When studying the rheology of well-entangled polymer melts, a non-stretching
approximation can be useful for both (1) studying flow behaviors that
are independent of $Z$ in the limit of $Z\gg1$ and (2) alleviating
computational challenges with fast-relaxing degrees of freedom, namely
chain stretching. However removing chain stretching altogether can
lead to pathological flow predictions including finite-time blow up,
and such pathological behaviors cannot generally be predicted a-priori.
Fortunately, for the RP model we have shown that a second-order expansion
of the non-stretching approximation preserves many of the advantages
of a non-stretching approximation while evidently avoiding its pathologies.
This is possible because the second-order expansion, which we call
nRP2, provides a more complete description of chain stretching.

\printbibliography

@inproceedings{premvzoe2003particle,
  title={Particle-based simulation of fluids},
  author={Prem{\v{z}}oe, Simon and Tasdizen, Tolga and Bigler, James and Lefohn, Aaron and Whitaker, Ross T},
  booktitle={Computer Graphics Forum},
  volume={22},
  number={3},
  pages={401--410},
  year={2003},
  organization={Wiley Online Library}
}

@inproceedings{muller2003particle,
  title={Particle-based fluid simulation for interactive applications},
  author={M{\"u}ller, Matthias and Charypar, David and Gross, Markus},
  booktitle={Proceedings of the 2003 ACM SIGGRAPH/Eurographics symposium on Computer animation},
  pages={154--159},
  year={2003},
  organization={Citeseer}
}

@article{kindt2007single,
  title={A single particle model to simulate the dynamics of entangled polymer melts},
  author={Kindt, P and Briels, Willem J},
  journal={The Journal of chemical physics},
  volume={127},
  number={13},
  year={2007},
  publisher={AIP Publishing}
}

@article{schieber2003full,
  title={A full-chain, temporary network model with sliplinks, chain-length fluctuations, chain connectivity and chain stretching},
  author={Schieber, Jay D and Neergaard, Jesper and Gupta, Sachin},
  journal={Journal of Rheology},
  volume={47},
  number={1},
  pages={213--233},
  year={2003},
  publisher={The Society of Rheology}
}

@article{khaliullin2010application,
  title={Application of the slip-link model to bidisperse systems},
  author={Khaliullin, Renat N and Schieber, Jay D},
  journal={Macromolecules},
  volume={43},
  number={14},
  pages={6202--6212},
  year={2010},
  publisher={ACS Publications}
}

@article{andreev2013approximations,
  title={Approximations of the discrete slip-link model and their effect on nonlinear rheology predictions},
  author={Andreev, Marat and Khaliullin, Renat N and Steenbakkers, Rudi JA and Schieber, Jay D},
  journal={Journal of Rheology},
  volume={57},
  number={2},
  pages={535--557},
  year={2013},
  publisher={AIP Publishing}
}

@article{giesekus1982simple,
  title={A simple constitutive equation for polymer fluids based on the concept of deformation-dependent tensorial mobility},
  author={Giesekus, Hanswalter},
  journal={Journal of Non-Newtonian Fluid Mechanics},
  volume={11},
  number={1-2},
  pages={69--109},
  year={1982},
  publisher={Elsevier}
}

@book{doi1988theory,
  title={The theory of polymer dynamics},
  author={Doi, Masao and Edwards, Sam F and Edwards, Samuel Frederick},
  volume={73},
  year={1988},
  publisher={oxford university press}
}

@article{likhtman2003simple,
  title={Simple constitutive equation for linear polymer melts derived from molecular theory: Rolie--Poly equation},
  author={Likhtman, Alexei E and Graham, Richard S},
  journal={Journal of Non-Newtonian Fluid Mechanics},
  volume={114},
  number={1},
  pages={1--12},
  year={2003},
  publisher={Elsevier}
}

@article{graham2003microscopic,
  title={Microscopic theory of linear, entangled polymer chains under rapid deformation including chain stretch and convective constraint release},
  author={Graham, Richard S and Likhtman, Alexei E and McLeish, Tom CB and Milner, Scott T},
  journal={Journal of Rheology},
  volume={47},
  number={5},
  pages={1171--1200},
  year={2003},
  publisher={The Society of Rheology}
}

@article{ianniruberto2001simple,
  title={A simple constitutive equation for entangled polymers with chain stretch},
  author={Ianniruberto, Giovanni and Marrucci, Giuseppe},
  journal={Journal of Rheology},
  volume={45},
  number={6},
  pages={1305--1318},
  year={2001},
  publisher={The Society of Rheology}
}

@article{kabanemi2009nonequilibrium,
  title={Nonequilibrium stretching dynamics of dilute and entangled linear polymers in extensional flow},
  author={Kabanemi, Kalonji K and H{\'e}tu, Jean-Fran{\c{c}}ois},
  journal={Journal of non-newtonian fluid mechanics},
  volume={160},
  number={2-3},
  pages={113--121},
  year={2009},
  publisher={Elsevier}
}

@article{boudara2019nonlinear,
  title={Nonlinear rheology of polydisperse blends of entangled linear polymers: Rolie-Double-Poly models},
  author={Boudara, Victor AH and Peterson, Joseph D and Leal, L Gary and Read, Daniel J},
  journal={Journal of Rheology},
  volume={63},
  number={1},
  pages={71--91},
  year={2019},
  publisher={The Society of Rheology}
}

@article{dolata2023thermodynamically,
  title={A thermodynamically consistent constitutive equation describing polymer disentanglement under flow},
  author={Dolata, Benjamin E and Olmsted, Peter D},
  journal={Journal of Rheology},
  volume={67},
  number={1},
  pages={269--292},
  year={2023},
  publisher={AIP Publishing}
}

@article{peterson2021predictions,
  title={Predictions for flow-induced scission in well-entangled living polymers: The living Rolie-Poly model},
  author={Peterson, Joseph D and Gary Leal, L},
  journal={Journal of Rheology},
  volume={65},
  number={5},
  pages={959--982},
  year={2021},
  publisher={The Society of Rheology}
}

@article{adams2011transient,
  title={Transient shear banding in entangled polymers: A study using the Rolie-Poly model},
  author={Adams, JM and Fielding, Suzanne M and Olmsted, Peter D},
  journal={Journal of Rheology},
  volume={55},
  number={5},
  pages={1007--1032},
  year={2011},
  publisher={AIP Publishing}
}

@article{likhtman2000microscopic,
  title={Microscopic theory for the fast flow of polymer melts},
  author={Likhtman, Alexei E and Milner, Scott T and McLeish, Tom CB},
  journal={Physical review letters},
  volume={85},
  number={21},
  pages={4550},
  year={2000},
  publisher={APS}
}

@article{read2008entangled,
  title={Entangled polymers: Constraint release, mean paths, and tube bending energy},
  author={Read, DJ and Jagannathan, K and Likhtman, AE},
  journal={Macromolecules},
  volume={41},
  number={18},
  pages={6843--6853},
  year={2008},
  publisher={ACS Publications}
}

@article{moorcroft2014shear,
  title={Shear banding in time-dependent flows of polymers and wormlike micelles},
  author={Moorcroft, Robyn L and Fielding, Suzanne M},
  journal={Journal of Rheology},
  volume={58},
  number={1},
  pages={103--147},
  year={2014},
  publisher={AIP Publishing}
}

@article{sharma2021onset,
  title={Onset of transient shear banding in viscoelastic shear start-up flows: Implications from linearized dynamics},
  author={Sharma, Shweta and Shankar, V and Joshi, Yogesh M},
  journal={Journal of Rheology},
  volume={65},
  number={6},
  pages={1391--1412},
  year={2021},
  publisher={AIP Publishing}
}

@article{chen2024derivation,
  title={Improving the functionality of non-stretching approximations},
  author={Chen, Vickie and Wang, Brandon and Peterson, Joseph D.},
  journal={Submitted to Journal of Rheology},
  publisher={American Institute of Physics}
}

@article{adams2009nonmonotonic,
  title={Nonmonotonic Models are Not Necessary to Obtain Shear Banding Phenomena<? format?> in Entangled Polymer Solutions},
  author={Adams, JM and Olmsted, Peter D},
  journal={Physical review letters},
  volume={102},
  number={6},
  pages={067801},
  year={2009},
  publisher={APS}
}

@article{carter2016shear,
  title={Shear banding in large amplitude oscillatory shear (LAOStrain and LAOStress) of polymers and wormlike micelles},
  author={Carter, Katherine A and Girkin, John M and Fielding, Suzanne M},
  journal={Journal of Rheology},
  volume={60},
  number={5},
  pages={883--904},
  year={2016},
  publisher={AIP Publishing}
}

@article{moorcroft2013criteria,
  title={Criteria for shear banding in time-dependent flows of complex fluids},
  author={Moorcroft, Robyn L and Fielding, Suzanne M},
  journal={Physical review letters},
  volume={110},
  number={8},
  pages={086001},
  year={2013},
  publisher={APS}
}

\end{document}